\documentclass[aps,prl,superscriptaddress,showpacs,twocolumn]{revtex4}
\usepackage{amsfonts}
\usepackage{amsmath}
\usepackage{amssymb}
\usepackage{graphicx}

\begin{document}

\title{Strong terahertz response in bilayer graphene nanoribbons}
\author{A. R. Wright}
\affiliation{School of Engineering Physics, University of
Wollongong, New South Wales 2552, Australia}
\author{J. C. Cao}
\affiliation{State Key Laboratory of Functional Materials for
Informatics,  Shanghai Institute of Microsystem and Information
Technology, Chinese Academy of Sciences, 865 Changning Road,
Shanghai 200050, P. R. China}
\author{C. Zhang$^{(a)}$}
\affiliation{School of Engineering Physics, University of
Wollongong, New South Wales 2552, Australia}

\begin{abstract}
We reveal that there exists a class of graphene structures (a
sub-class of bilayer graphene nanoribbons) which has unusually
strong optical response in the terahertz (THz) and far infrared
(FIR) regime. The peak conductance of terahertz/FIR active bilayer
ribbons is around two orders of magnitude higher than the
universal conductance of $e^2/4\hbar$ observed in graphene sheets.
The criterion for the terahertz/FIR active sub-class is a bilayer
graphene nanoribbon with one-dimensional massless Dirac Fermion
energy dispersion near the $\Gamma$ point. Our results overcome a
significant obstacle that hinders potential application of
graphene in electronics and photonics.
\end{abstract}

\pacs{73.50.Mx, 78.67.-n, 81.05.Uw} \maketitle

In recent years, graphene has attracted a great deal of interest\cite%
{novo1,novo2,zhang1,berg}. New physics have been predicted and
observed, such as
electron-hole symmetry and half-integer quantum Hall effect\cite%
{novo2,zhang1}, finite conductivity at zero charge-carrier
concentration\cite{novo2}, and the strong suppression of weak
localization\cite{suzuura,morozov,khveshchenko}. Bilayer graphene
(BLG) has also attracted considerable attention recently, with
seminal experimental and theoretical work being carried out
\cite{novo1,McCann}. Bilayer calculations use interlayer coupling
constants based on the Slonczewksi-Weiss-Mclure model \cite{S-W,
McClure}. By further confining the electrons in the graphene
plane, one can obtain one-dimensional structures which we refer to
as graphene nanoribbons (GNRs)\cite{young}. It has been shown that
GNRs with zigzag edges can have finite magnetization with either
ferromagnetic order or antiferromagnetic order\cite{louise,yan}.
These properties promise building blocks for technological
applications in molecular electronic and optoelectronic devices.

The optical properties of graphene systems is a topic of
considerable interest. In particular, the minimal conductivity of
single layer graphene (SLG) within the Dirac regime is a much
celebrated result which was calculated theoretically long before
graphene's fabrication in 2003 \cite{early_universal}. The optical
conductivity of SLG outside the low energy Dirac regime has been
calculated theoretically\cite{Chao-Ma, peres_OC}. The conductance
of bilayer graphene has also been calculated \cite{trig_effect},
as has the conductance of various GNRs \cite{Chao-GNRsOCBField}.
All of this research has shown that the optical responses of
graphene and graphene nanoribbons are extremely weak. In the EM
frequency band from terahertz to visible, the absorption
coefficient for these systems is generally less than
3\%.\cite{gus,kuz,nair}. While the lateral confinement and the
edge states can lead to the ferromagnetic and antiferromagnetic
order in GNRs\cite{louise}, the optical response of all ribbons
remains very weak. There are two fundamental reasons for this: (1)
the density of states vanishes near the Fermi energy, and (2) the
interband transition amplitude is small. Because of this,
potential application of graphene structures in optoelectronics
and photonics is severely limited. To date, these obstacles have
remained.

However we are now able to demonstrate that there exists a
sub-class of bilayer graphene nanoribbons (BLGNRs) which have an
unusually strong optical conductance in the terahertz (THz) to far
infrared (FIR) regime. The height of the conductance peak is close
to two orders of magnitude greater than the universal conductance
of graphene sheets. We found that this sub-class of BLGNRs can be
either armchair or chiral, but their energy dispersion near the
$\Gamma$ point must be that of a one-dimensional massless Dirac
Fermion. This sub-class of graphene structures are the first
systems to show such a strong optical response in the absence of
any external field in the important frequency band of THz and FIR.

We first construct a single layer GNR (SLGNR) following the
convention of Ezawa \cite{Ezawa}, i.e., a ribbon is specified by
two indices $p$ and $q$. We begin by placing $m = p+q$ hexagons
next to each other with flat edges touching. On top of this layer,
we place an identical layer, offset by $q$ hexagons. Continuing
this in both directions we can construct a GNR of arbitrary
chirality. Within this model $q = 0$ corresponds to a zig-zag (ZZ)
edged ribbon, and $q = 1$ corresponds to an armchair (AC) edged
ribbon. $q>1$ corresponds to arbitrary chirality, defined by an
angle $\theta = \tan^{-1}(\sqrt{3}/(2q+1))$. The number of atoms
in the unit cell is given by $N_u = 4q + 2p + 2$. The second layer
is now constructed by assuming the standard $A-B$ `Bernal'
stacking, along the vertical C-C vector. The net effect is simply
shifting the entire second layer up (or down) by an amount $C-C =
1.42\AA$. A typical BLGNR is shown in Figure 1.

\begin{figure}[tbp]
\centering\includegraphics[width=8cm,height=3.5cm]{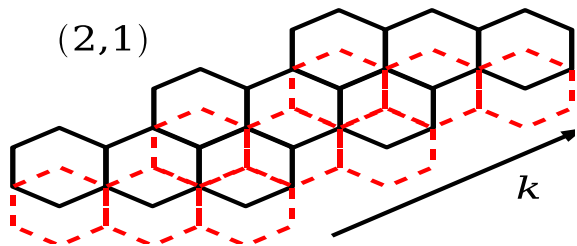}
\caption{The construction of a bilayer graphene nanoribbon. The
second layer is identical to the first, but shifted along a C-C
vector.}
\end{figure}

The intralayer coupling is calculated using the tight binding
formalism where $t \approx 3eV$  is the nearest neighbour hopping
integral. The edge effects of GNRs are incorporated into the tight
binding formalism by setting the overlap integral to zero for
hopping between edge sites and their neighbours which are off the
edge of the ribbon. In this way, the edge states are incorporated
into the full electronic properties. The interlayer coupling is
restricted to the dominant coupling term $\gamma = 0.36eV$ which
occurs only between $A$ and $B$ sites which sit directly one on
top of the other. As an example the Hamiltonian matrix for a (2,1)
BLGNR is given by
\begin{equation}
H_{(2,1)} = \left(
\begin{array}{cc}
H_{\mathrm{intra}} & H_{\mathrm{inter}}\\ H_{\mathrm{inter}}^* &
H_{\mathrm{intra}}
\end{array}
\right)
\end{equation}
Where the intralayer terms are given by

\begin{equation}
H_{\mathrm{intra}} = \left(
\begin{array}{cc}
0 & h\\
h^* & 0
\end{array}
\right)
\end{equation}
And
\begin{equation}
h = \left(
\begin{array}{ccccc}
h_1 & h_2 & 0 & 0 & 0 \\ h_1^* &
h_1 & h_2 & 0 & 0\\ 0 & h_1^* &
h_1 & 0 &  h_2\\ 0 & 0 & h_1^* &
h_2 & h_1\\ 0 & 0 & 0 & h_1 &
h_1^*
\end{array}
\right)
\end{equation}
Where $h_1 = e^{ikb\sqrt{3}/2}$, and $h_2 = e^{ikb}$. The
interlayer coupling matrix is given by

\begin{equation}
H_{\mathrm{inter}} = \left(
\begin{array}{cc}
0 & h'\\
h^{T'} & 0
\end{array}
\right)
\end{equation}
Where

\begin{equation}
h' = \left(
\begin{array}{ccccc}
0 & \gamma & 0 & 0 & 0\\
0 & 0 & \gamma & 0 & 0\\
0 & 0 & 0 & 0 & \gamma\\
0 & 0 & 0 & \gamma & 0\\
0 & 0 & 0 & 0 & 0\\
\end{array}
\right).
\end{equation}

The band structures for ZZ-BLGNRs are shown in Figure 2(a). They
differ from the single layer case as the interlayer coupling
produces a second subband corresponding to each single layer
subband offset by an amount $~\gamma$ for most of the Brillouin
Zone. Near the Dirac-like points however, the intersubband gap
goes to zero for the low energy bands which approach zero energy,
and the higher energy bands all approach a degenerate point
slightly away from the Dirac-like points. The edge states cause
the plateau of the zigzag dispersion at low energies, causing an
extended region of zero band-gap near the zone boundary.

\begin{figure}[tbp]
\centering\includegraphics[width=8cm]{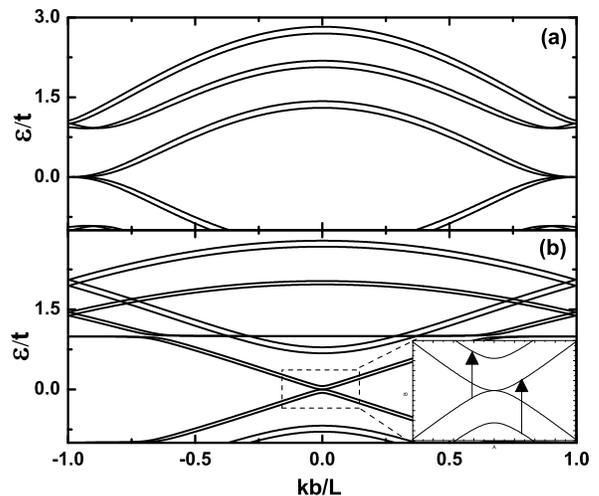} \caption{The
bilayer graphene nanoribbon electronic dispersion curves differ
from the single layer  ones as each single layer subband has a
complementary subband which generally differs in energy by
$<\gamma$. The zigzag case is given in (a), and the Dirac armchair
case in (b). The inset to (b) shows the two sets of symmetric
bands near the Dirac point. The arrow represents transitions
between non-symmetric bands which is the dominant transition at
low energies that leads to the unusually high optical conductance
observed in figure 3(b).}
\end{figure}

The band structures of AC-BLGNRs shown in figure 2(b) behave much
like 2D bilayer graphene. The linear dispersion at the Dirac
points becomes curved, and a second subband appears. Each subband
is thus paired. But unlike the ZZ-BLGNR case, the intersubband gap
does not go to zero near the Dirac points, but remains at least
$\gamma/2$ separated from it's pair. As the width increases, this
gap approaches  zero as the band pairs converge. This is strictly
an edge effect, and in the 2D limit a band gap of $\gamma$
re-emerges.

We shall show below that the oscillator strength of the interband
transition is strongly dependent on the properties of the energy
dispersion at the zero-gap point. We shall show below that for
ZZ-BLGNR, where the zero gap position is at the K point and the
two low energy dispersions are close to parabolic, the oscillator
strength at low energy is very small. On the other hand, for
AC-BLGNR, the zero gap position is at the $\Gamma$ point.
Furthermore, the two low energy dispersions are very close to
linear (or a one-dimensional massless Dirac Fermion). In this case
the interband transition between the bands (transitions between
the non-symmetric bands shown by arrows in the inset of figure
2(b)) is extremely strong.

\begin{figure}[tbp]
\centering\includegraphics[width=8cm]{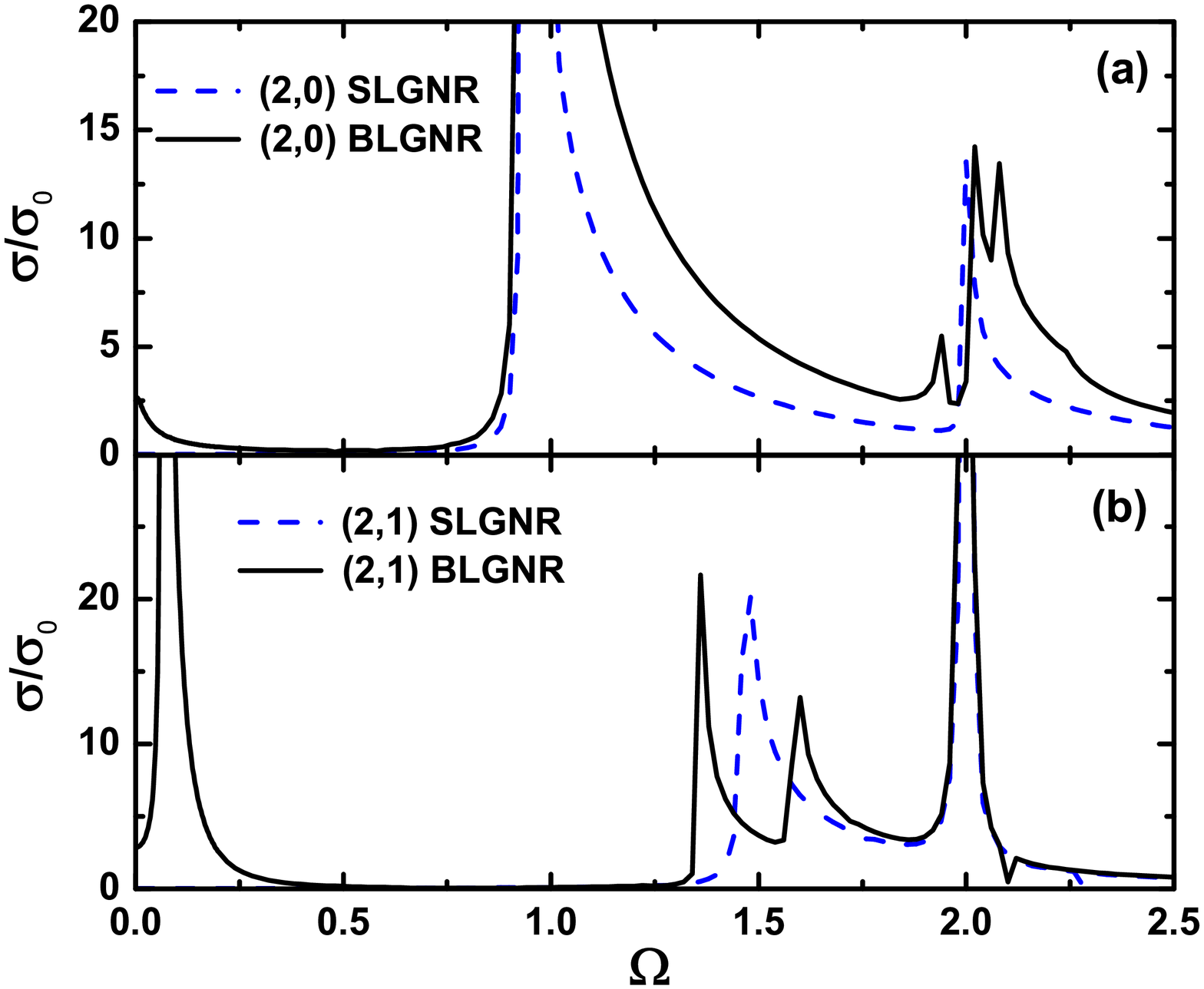} \caption{The
optical conductivity for the ZZ-BLGNR (a) and the AC-BLGNR (b).
The low energy activity in the bilayer ribbons is particularly
significant, especially in the Dirac AC-BLGNR where the optical
conductivity is approximately $80\sigma_0$.}
\end{figure}

The optical conductivity is calculated using the Kubo formula
given by

\begin{equation}
\sigma(\omega) =
\frac{1}{\omega}\int_0^\infty\mathrm{d}te^{i\omega t}
\langle[J(t),J(0)]\rangle
\end{equation}
Where $J$, the current operator is given by $\partial H/ \partial
y$. We define the dimensionless photon frequency $\Omega =
\hbar\omega/t$,  and normalize our results to the single layer
universal conductivity given by $\sigma_0 = e^2/4\hbar$. We
determine the dependence of the optical conductance on the ribbon
widths and chiralities. In figure 3(a) we show that the optical
conductivity for the ZZ-BLGNR exhibits a spike centered on zero
energy. This spike occurs here because both low energy subbands
approach zero energy at the Dirac-like points. In the SLGNR case,
the velocity operator approaches a constant, which makes
inter-symmetric-subband transitions forbidden. This is no longer
the case in bilayer ribbons, and there is also now the possibility
of low energy inter-non-symmetric-subband transitions.

Over the full energy spectrum, we see that some of the resonant
peaks in the single layer optical conductivity spectrum have split
into three peaks. This will not generally be the case. Most peaks
will split into two as will be seen in the armchair case. However,
near $\Omega = 1$, the subbands create a linear
Dirac-like band structure with features similar to the Dirac point
in the proceeding armchair case, as well as those observed in 2D
bilayer graphene. This means that there are three possible energy
transitions with high density of states. The central, primary peak
corresponds to the original SLGNR peak, and the two secondary
peaks, one below, and and one above the original by an amount
$\Omega = \gamma$, correspond to the new curved subbands
which don't quite touch the degenerate point from the single layer
case.

The optical conductance of armchair BLGNRs is shown in figure
3(b), In the AC case it peaks sharply at $\gamma/2$ and trails off
because of the curvature of the band structure. In the armchair
case, a peak is still observed as $\Omega \rightarrow 0$, but the
peak at $\gamma/2$ is about 2 orders of magnitude stronger. This
peak corresponds to vertical transitions between non-symmetric
subbands which are far more probable than those between symmetric
ones (see fig. 5 and subsequent discussion). This single low
energy peak is larger than every other peak across the spectrum.
As the width of the ribbons increases, the strength of this peak
decreases and the width of it broadens. In the case of an
infinitely wide ribbon, the strong optical response is lost,
settling at $\approx 4\sigma_0$ at $\Omega = \gamma$. This
reflects the peculiar edge dependence of the bilayer ribbon
system.

\begin{figure}[tbp]
\centering\includegraphics[width=8cm]{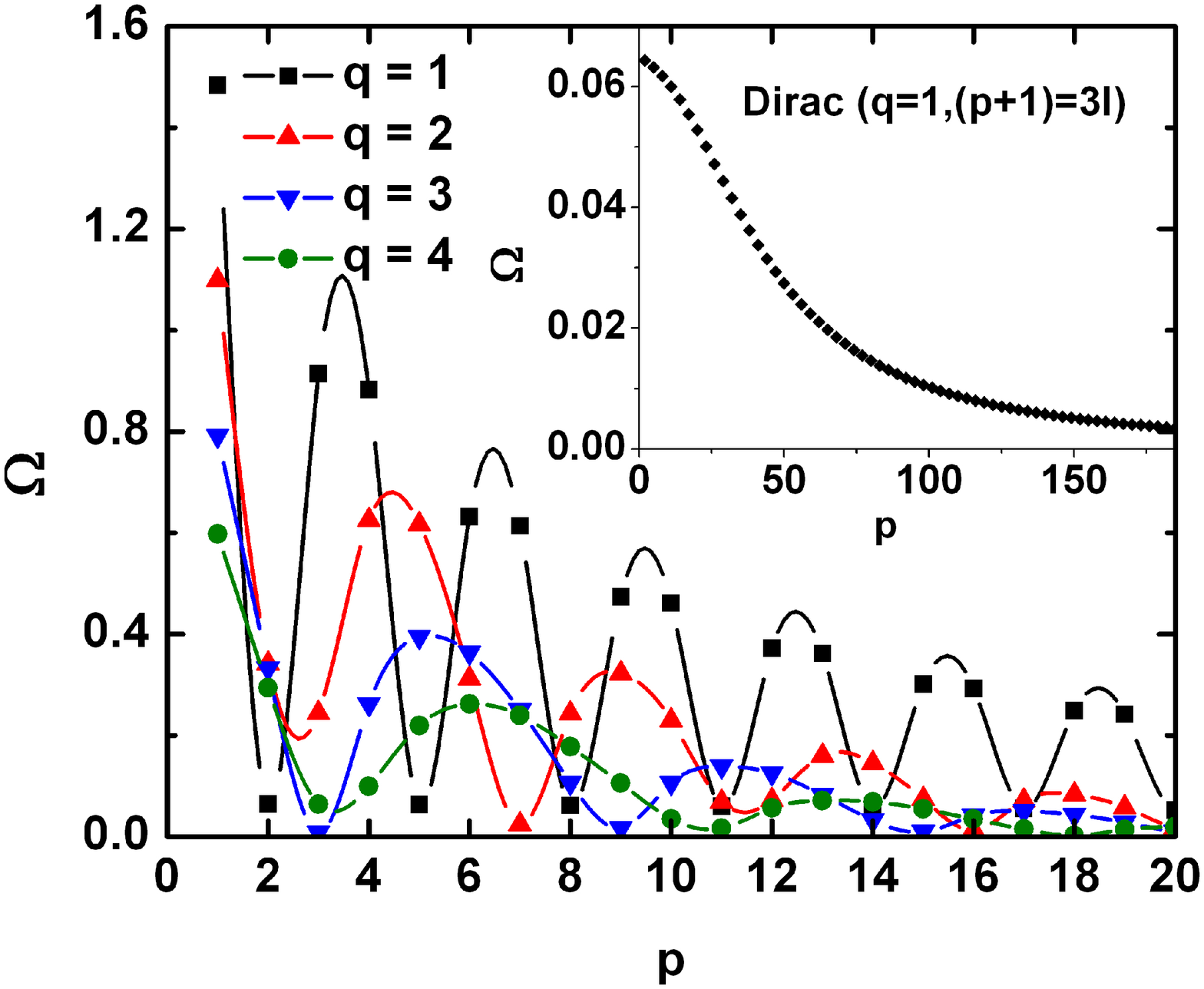} \caption{The width
dependence of the energy of the low-energy peak for BLGNRs with $q
= [1,4]$. The strongest peaks occur in the lowest energy gaps in
the Dirac armchair ribbons. The inset shows the  width dependence
of the band gap for Dirac BLGNRs with strong low energy optical
response. This gap eventually disappears, but in the 2D limit with
no edges, it re-emerges at $\gamma$.}
\end{figure}

At higher energies, the
single peaks observed in AC-SLGNRs generally have split into two,
and  are separated by an amount $\approx 2\gamma$. This
corresponds to two sets of symmetric transitions, the
non-symmetric transitions being largely suppressed.

For $q>1$ BLGNRs, the band gap between the two lowest energy
symmetric bands varies from zero to $~1eV$. Similarly the
non-Dirac AC-BLGNRs (ie $(p+q)/3 \notin I$), have varying
band-gaps for the lowest energy subbands.

\begin{figure}[tbp]
\centering\includegraphics[width=8cm]{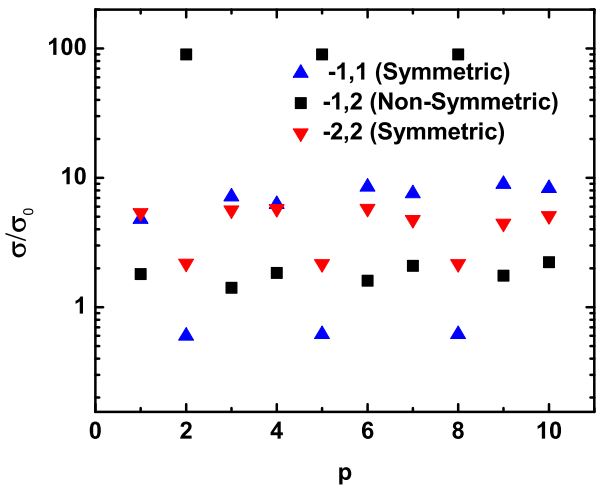} \caption{The p
dependence of the low energy peaks for a (p,1) BLGNR. In Dirac
ribbons the symmetric transitions are suppressed and the
non-symmetric ones dominate. For non-Dirac ribbons the opposite is
the case.}
\end{figure}

For a given type of BLGNR, the conductance peak position can be
tuned with the ribbon width. Figure 4 shows the width dependence
of the peak position in the THz/FIR regime. The peak position
oscillates with the ribbon width. The amplitude of the oscillation
is of the same order of magnitude as the average peak position,
indicating a large range for tuning the resonance peak. The period
of the oscillation increases with the chirality (q). The inset of
figure 4 shows the width dependence of the energy gap for Dirac
BLGNRs. This gap decreases as width increases making the location
of the optical peak strongly width dependent. In the limit of
infinite width this edge determined gap approaches zero, and the
2D bilayer band structure sees the emergence of a gap of $\gamma$.
The optical response in this case however is not nearly so large,
being $\approx 4\sigma_0$

Figure 5 shows the width dependence of the magnitude of the low
energy peak for the $(p,1)$ BLGNR. For $(p + 1)/3 \in I$, the
non-symmetric matrix element dominates, causing the single low
energy peak. The low-frequency peak conductance for this class of
BLGNRs is unusually strong having a value of approximately
$80\sigma_0$, much stronger than the universal conductance of
graphene sheets.\cite{nair} When the Dirac condition is not met
however (ie $(p + 1)/3 \notin I$), the symmetric matrix elements
dominate, and the non-symmetric matrix  elements are greatly
suppressed, leading to a much weaker response to the low energy
spectrum.

\begin{figure}[tbp]
\centering\includegraphics[width=8cm]{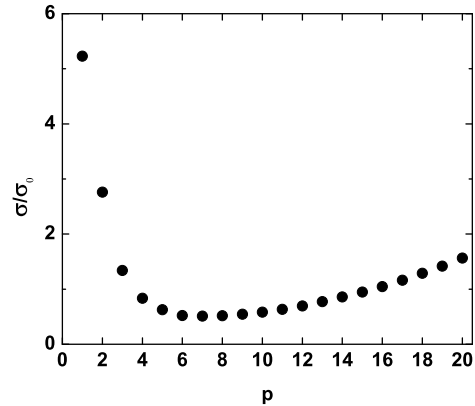} \caption{The width
dependence of the magnitude of the low energy peak  for the
ZZ-BLGNRs decreases quickly with increasing width, and then
increases steadily for $p > 6$.}
\end{figure}

The width dependence of the strength of the low energy peak for
ZZ-BLGNRs is given in Figure 6. For very narrow ribbons with $p<6$
the peak quickly decreases in magnitude at a decreasing rate. For
$p>6$ however, the peak magnitude increases steadily reflecting
the low energy subband shape. As the width increases, the low
energy subbands remain lower, which increases the DOS, allowing
more transitions between subbands. For very narrow width ZZ-BLGNRs
however, the curvature in the subbands is so high that the
velocity operator allows strong coupling between the subbands,
which makes the low energy magnitude very strong.

In summary, we have shown that the interplay of ribbon's chirality
and the inter-ribbon coupling can lead to significant enhancement
in optical response. We have identified a sub-class of BLGNRs
where the inter-ribbon coupling causes a finite band gap in the
energy minimum (maximum) and induces strong inter-subband
transitions. The distinct feature of this sub-class of BLGNRs is
that they have a one dimensional massless Dirac Fermion dispersion
near the $\Gamma$ point. The peak conductance of this class of
BLGNRs has a very large value of as much as $80\sigma_0$, making
them a class of materials for unique applications in
optoelectronics. The simple picture behind this phenomenon is that
the density of states for the 1D massless Dirac Fermions remains
finite at zero energy, whereas that for the 2D massless Dirac
Fermions in a graphene sheet vanishes. The peculiar role of the
edge states has also been shown to contribute as the 2D limit of
infinite width shows markedly decreased response in this frequency
regime.

These results reported here open a gateway to the creation of
graphene-based low energy photon devices.  The ribbon width and
chirality selection for various applications is crucial, as the
optical responses of various ribbons change dramatically when
these properties are varied.

\bigskip
\noindent {\it Acknowledgement}--This work is supported in
part by the Australian Research Council.

\bigskip
\noindent{$^{a)}$Electronic mail: czhang@uow.edu.au}\\

\end{document}